\documentstyle[aps,preprint,epsf,array]{revtex}

\begin{document}
\title{\bigskip Thermal Conductance for Single Wall Carbon Nanotubes}
\author{Qingrong Zheng$^{a,c}$, Gang Su$^{a,*}$, Jian Wang$^{b}$ and Hong Guo$^{c}$}
\address{$^{a}$Department of Physics, The Graduate School of the Chinese \\
Academy of Sciences, P.O. Box 3908, Beijing 100039, China}
\address{$^b$Department of Physics, The University of Hong Kong,\\
Pokfulam Road, Hong Kong}
\address{$^c$Centre for the Physics of Materials, Department of Physics,\\
McGill University, Montreal, Quebec, Canada H3A 2T8}
\maketitle

\begin{abstract}
We report a theoretical analysis of the phonon thermal conductance, $\kappa
(T)$, for single wall carbon nanotubes (SWCN). In a range of low temperatues
up to 100$K,$ $\kappa (T)$ of perfect SWCN is found to increase with
temperature, approximately, in a parabolic fashion. This is qualitatively
consistent with recent experimental measurements where the tube-tube
interactions are negligibly weak. When the carbon-carbon bond length is
slightly varied, $\kappa (T)$ is found to be qualitatively unaltered which
implies that the anharmonic effect does not change the qualitative behavior
of $\kappa (T)$.

\end{abstract}

\newpage

\section{Introduction}

Since its original discovery, carbon nanotubes\-\cite{ref1} have received a
great deal of attention due to fundamental physical interest on nano-scale
systems, as well as due to nanotubes$^{\prime }$ potential for useful
industrial applications \cite{ref2}. A very important recent advance has
been the fabrication of high-purity crystalline bundles of nearly
mono-disperse single wall carbon nanotubes\-\cite{ref3,ref4,ref5}. This
allows better experimental control and produces accurate data. It also
provides opportunities and points to new directions for theoretical analysis
of SWNT. So far, the electronic and mechanical properties of carbon
nanotubes have been extensively investigated while there also exist several
new measurements of thermal properties of these systems\cite
{ref6,ref7,ref8,ref9}. The purpose of this work is to present our
theoretical and numerical analysis of thermal conductance of SWNT.

Of particular interest to this work is the recent experimental measurements
of thermal conductance of nanotubes $\kappa (T)$\cite{ref6}. This quantity
describes the thermal current induced by a temperature gradient, and \ we
will analyze a two-terminal measurement of it. Hence, maintaining
temperatures at the left and right lead to be $T_{L},T_{R}$ respectively, $%
\kappa (T)$ is defined as $\kappa (T)\equiv \dot{Q}/\nabla T$, where $\dot{Q}
$ is the thermal current flowing through the nanotube. The experimentally
measured thermal conductance of carbon nanotubes indicates that the most
essential contribution comes from phonons\cite{ref6,ref7}. The data of Ref. 
\cite{ref6} on aligned multiwall nanotubes suggested that at low temperature
up to $\sim 120K,$ $\kappa (T)$ can be well fit by quadratic form in $T$, 
{\it i.e. }$\kappa (T)\sim T^{2}$. \ On the other hand, the measurements\cite
{ref8} of specific heat $C(T)$, which is proportional to $\kappa (T)$, also
showed a power law dependence, however with a power larger than unity but
slightly less than two\cite{ref8}. Finally, measurements on crystalline
ropes of SWNT indicates\cite{ref7} a linear temperature dependence up to $%
30K $ and an upward bend near $30K$ on the $\kappa (T)$ versus $T$ curve 
\cite{ref7}. How to interpret these experimental observations presents a
challenge to theory. Indeed, although there exist a vast literature on
theoretical analysis of thermal transport of both electrons and phonons in
the context of mesoscopic physics\cite
{ref11,ref12,ref13,ref14,ref15,ref16,ref17,ref18,ref19}, direct
investigation on carbon nanotubes is rare\cite{ref8,ref10,berber}.

The precise temperature power law in thermal conductance of nanotubes is
expected to depend on, among other things, the detailed phonon dispersion.
In this work we will calculate the dispersion for several SWNT and
investigate its consequence. We neglect electrons as their contribution to $%
\kappa (T)$ of nanotubes can only be observed at temperature less than $1K$ 
\cite{ref10}. Our analysis of SWNT is based on the Tersoff-Brenner$\prime $s
empirical potential\cite{ref20,ref21} for carbon to calculate the phonon
dispersion. We then apply a Landauer-Buttiker-type formalism to compute the
lattice thermal conductance in various SWNTs, for which the thermal
transmission coefficient is assumed to take a Breit-Wigner form. The
predicted $\kappa (T)$ shows a quadratic form in temperature for both zigzag
and armchair SWNT. Using one experimental data to fix an overall shift of $%
\kappa (T)$, our predicted $\kappa (T)$ is consistent with the experimental
data of Ref.\cite{ref6} as the tube-tube interactions are so weak that we
could compare our calculations for single wall carbon tubes with the
experimental results on multiwall tubes. As pointed out in Ref. \cite{ref7},
since the tube-tube interactions are also weak, the measured $\kappa (T)$ is
linear below $30K$ and shows an upward bend slightly near $30K$, i.e. the
curve appears to be also parabolic-like from $8K$ to $100K$. When the
carbon-carbon bond length is slightly varied, our results indicate that $%
\kappa (T)$ does not change qualitatively, implying that the anharmonic
effect does not alter its qualitative behavior.

The rest of this paper is organized as follows. In the next section we
outline the analysis of phonon dispersion and $\kappa (T)$. Section III
presents the numerical data while a brief summary is given in the last
section.

\section{Lattice Thermal Conductance}

We start by writing down a multi-probe formula for thermal current
transmission in the same spirit as the familiar Landauer-Buttiker formula 
\cite{ref22} for electron transport. As discussed above we neglect
contributions from electrons and deal with the lattice vibration as a phonon
gas. The thermal current is therefore driven by a temperature gradient and
carried by phonons from the probe to another. The thermal current can be
written as

$$
\dot{Q}^{\alpha }=\frac{1}{h}\int dE\text{ }E\text{ }\sum_{\beta }f_{\beta
}(E)\text{ }A_{\alpha \beta }(E)\eqno(1) 
$$
where $\alpha ,\beta $ label the leads, $h$ is the Planck constant, $E$ is
the phonon energy, and $f_{\beta }(E)=1/(e^{\frac{E}{k_{B}T_{\beta }}}-1),$
is the phonon distribution function with temperature $T_{\beta }$ at lead $%
\beta $ . The thermal transmission function $A_{\alpha \beta }(E)$ is
defined as 
$$
A_{\alpha \beta }(E)=\delta _{\alpha \beta }-S_{\alpha \beta }^{\dagger
}(E)S_{\alpha \beta }(E),\eqno(2) 
$$
where $S_{\alpha \beta }$ is the scattering matrix for phonons. Eqs.(1,2)
are similar to the Landauer-Buttiker formula \cite{ref22} for electron
transport. Let the energy of the $s$-th branch of phonon be $\hbar \omega
_{s}$. The thermal current can be rewritten as 
$$
\dot{Q}^{\alpha }=\frac{1}{2\pi }\sum_{s}\int d\omega _{s}\text{ }\hbar
\omega _{s}\text{ }\sum_{\beta }f_{\beta }(\hbar \omega _{s})\text{ }%
A_{\alpha \beta }(\hbar \omega _{s})\eqno(3) 
$$
or 
$$
\dot{Q}^{\alpha }=\frac{1}{2\pi }\sum_{s}\int_{0}^{\infty }d{\bf q}\cdot 
{\bf v}_{s}({\bf q})\text{ }\hbar \omega _{s}({\bf q})\text{ }\sum_{\beta
}f_{\beta }({\bf q})\text{ }A_{\alpha \beta }({\bf q}),\eqno(4) 
$$
where ${\bf v}_{s}({\bf q})=\nabla _{{\bf q}}\omega _{s}({\bf q})$ is the
group velocity.

For the special case of two probe measurement in one dimension, one can
easily confirm that Eq.(4) is reduced to the form which has been used to
study the quantized thermal conductance of dielectric quantum wires \cite
{ref23}. We have 
$$
\dot{Q}=\sum_{s}\int_{0}^{\infty }\frac{dk}{2\pi }\hbar \omega
_{s}(k)v_{s}(k)(f_{R}-f_{L})\zeta _{s}(k),\eqno(5) 
$$
where $\omega _{s}(k)$ and $v_{s}(k)=\frac{\partial \omega _{s}(k)}{\partial
k}$ are the frequency and the velocity of normal mode $s$ of the wire with
wave vector $k$, respectively, $\zeta _{s}(k)$ is the phonon transmission
probability through the wire, and $f_{\alpha }$ is the phonon distribution
function at the right or left lead indicated by $\alpha $ $(=R,L)$. We refer
interested readers to Refs.\cite{ref24,ref25} for more details about the
present formulation. For the SWNT the thermal conductance, $\kappa =\dot{Q}%
/\nabla T$, can be numerically computed through the following form 
$$
\kappa =\frac{\Delta L}{2\pi }\sum_{s}\int_{\omega _{s}(0)}^{\omega _{\max
}}d\omega \hbar \omega (f_{R}(\omega )-f_{L}(\omega ))\zeta _{s}(\omega
)/\Delta T.\eqno(6) 
$$
where $\Delta T=T_{R}-T_{L}$ is the temperature difference between the two
leads, $\Delta L$ is the length of the SWNT, and $\omega _{\max }$ is the
maximum of phonon modes in SWCNs, which was investigated elsewhere\cite
{ref25-1}.

To proceed, it is clear that the phonon frequencies of SWNT at $\Gamma $
point and the transmission function $\zeta _{s}(\omega )$ are required.
There are a number of methods which can be applied to obtain the phonon
frequencies, and we will present our calculation of this quantity in the
next section. On the other hand, it is rather complicated to accurately
determine the phonon transmission function. For a perfect infinite harmonic
lattice the thermal conductivity diverges because every mode transmits
perfectly. 
The contacts play an important role as in case of electron transport. Even
for perfect phonon transmission, the thermal conductance is finite due to
the contact resistance. 
Defects, boundary scatterings, phonon-phonon interactions and other
scattering mechanisms in the conductor inevitably introduce thermal
resistances. The transmission function $\zeta _{s}(\omega )$ is contributed
by lattice imperfection and conductor boundary, and the anharmonic effect
due to interactions can be included in terms of the renormalized
temperature-dependent dispersion relation. To obtain $\zeta _{s}(\omega )$
for SWNT one should in principle solve the scattering of a phonon wave by
the atoms of the nanotube which is a nontrivial problem not solved so far.
However if the length of a conductor is much larger than its cross-section
size, $\zeta _{s}(\omega )$ can be approximately obtained by solving linear
elastic equations assuming the conductor to be an uniform elastic medium\cite
{ref23}. In 1D, the continuum mechanics results indicate\cite{ref23} that $%
\zeta _{s}(\omega )$ is dominated by sharp resonances reminiscent of a
resonance transmission of phonon modes: $\zeta _{s}(\omega )=1$ at resonant
energies and sharply becomes much smaller for other energies. Guided by this
observation and for simplicity of analysis, we shall investigate the
consequence of such a resonance transmission of phonon modes for carbon
nanotubes by parameterizing $\zeta _{s}$ into a Breit-Wigner form,

$$
\zeta (E)=\frac{D^{2}}{(E-E_{0})^{2}+D^{2}},\eqno(7) 
$$
where $D$ and $E_{0}$ are the width of the resonance and position of a
resonance. Both of these parameters are closely related to the effects of
defects and boundary scatterings. Furthermore, instead of directly
considering \ anharmonic effects, we will investigate its consequence by
simply varying the carbon-carbon bond length to observe how much the thermal
conductance can vary.

\section{Lattice Vibrations}

To calculate the phonon spectra $\omega _{s}$ at the $\Gamma $ point for
armchair and zigzag carbon nanotubes, we shall make use of the
Tersoff-Brenner empirical potentia\cite{ref20,ref21} for carbon. Although
this empirical potential only gives an approximation to the vibration
frequency of nanotubes, it does capture the basic and qualitative
characteristics of their phonon spectra. Tersoff-Brenner$^{\prime }$s
potential is also numerically efficient to allow investigations of large
systems which is required for our study. The parameters for this potential
have been determined before and there are two sets of them which have been
widely used \cite{ref21} for structure analysis of carbon: 
\[
\begin{array}{ccccccccccc}
& R_{ij}^{(e)}\text{\r{A}} & D_{ij}^{(e)}\text{eV} & \beta _{ij}/\text{\r{A}}
& S_{ij} & \delta _{i} & R_{ij}^{(1)}\text{\r{A}} & R_{ij}^{(2)}\text{\r{A}}
& \alpha _{0} & c_{0} & d_{0} \\ \hline
1^{st} & 1.315 & 6.325 & 1.5 & 1.29 & 0.80469 & 1.7 & 2.0 & 0.011304 & 19.0
& 2.5 \\ 
2^{nd} & 1.39 & 6.0 & 2.1 & 1.22 & 0.5 & 1.7 & 2.0 & 2.0813\times 10^{-4} & 
330 & 3.5 \\ \hline
\end{array}
\]
The meaning of the symbols refers to Ref.\cite{ref21}. 
We have checked that for SWNT the second set of Brenner$\prime $s parameters
gave results closer to those obtained by using Tersoff parameter\cite{ref20}%
. We shall use in our calculations the second set values shown in the table.
In addition, experimental measurements have shown that phonon modes of SWNT
have frequencies up to $1700cm^{-1}$. We have numerically confirmed that if
the equilibrium distance parameter $R_{ij}^{(e)}=1.33$\r{A}, which is a
value in between the first and second Brenners$\prime $s parameter sets (see
table), the obtained vibration frequencies at $\Gamma $ point is consistent
with experimental data\cite{ref26}. \ On the other hand, all the parameter
sets give similar total binding energy. For instance, for the armchair (4,4)
nanotube, the binding energy per atom is found to be -6.8920, -7.0881,
-6.9993 and -7.0456 eV using Tersoff$^{\prime }$s potential, the first
Brenner$^{\prime }$s potential, the second Brenner$^{\prime }$s potential
and our modified second Brenner$^{\prime }$s potential, respectively. Fig. 1
presents a comparison between these empirical potentials. In all subsequent
numerical analysis we shall utilize the revised value of $R_{ij}^{(e)}$.

The lattice vibration frequencies of the SWCN at $\Gamma $ point can be
obtained by diagonalizing the dynamical matrix given by 
$$
D_{\alpha \beta }(k,lk^{\prime })=\frac{1}{M}\sum_{l}\Phi _{\alpha \beta
}\left( 0k,lk^{\prime }\right) e^{i{\bf q\cdot (R}(0k){\bf -R}(lk^{\prime }%
{\bf )}},\eqno(10) 
$$
where ${\bf q}$ is the wave vector, the perturbation of $R_{\alpha }(lk)$ is
taken to be $0.001$\r{A}, and 
\[
\Phi _{\alpha \beta }\left( lk,l^{\prime }k^{\prime }\right) =\frac{\partial
^{2}E_{b}}{\partial R_{\alpha }\left( lk\right) \partial R_{\beta }\left(
l^{\prime }k^{\prime }\right) }. 
\]
Once the phonon modes are obtained, we are able to calculate thermal
conductance by Eq. (6). Here we would like to point out that although there
have been a number of studies dedicated to the lattice vibrations of carbon
nanotubes so far\cite{ref2}, it is not suitable for us to adopt directly the
phonon data in literature. This is because in our situation we need
frequency data for different sizes of carbon nanotubes as well as the data
of frequencies when the carbon-carbon bond length is simply varied in order
to retrieve the anharmonic effects. In principle, we can use ab initio
molecular dynamics method to get the frequencies if the system is not too
big. Since in the present case the system contains hundreds of atoms, it is
not easy to obtain useful data on the basis of ab initio simulations, and
for a practical purpose it is better to invoke an empirical potential to
compute the phonon spectra of SWNCs with larger sizes.

\section{Results}

As discussed in the introduction, we are investigating long SWCNs such that
the left end is at temperature $T_{L}$ and the right end is at temperature $%
T_{R}.$ Let us first consider the basic characteristic of $\kappa (T)$. At
very low temperatures where only the massless modes (i.e. $\omega _{s}(0)=0$%
) are relevant, $\kappa /T$ is simply a constant independent of temperature.
At slightly higher temperature, the high frequency modes set in because a
SWNT is a quasi-one dimensional system in which each unit cell contains so
many atoms that there are many modes with nonzero cutoff frequencies.
Therefore $\kappa /T$ will deviate from the constant behavior when
temperature is increased.

The inset of Fig. 2 shows the temperature dependence of thermal conductance
for the armchair (11,11) and the zigzag (11,0) carbon nanotubes when the
transmission is perfect, {\it i.e.,} by setting $\varsigma _{s}=1$ for all
the phonon modes. For these cases our results indicate that $\kappa (T)$
increases, approximately, in a parabolic fashion as a function of
temperature. We found that anharmonic effect has only a slight influence on
the temperature dependence of $\kappa (T)$, and this is shown by calculating 
$\kappa (T)$ for different values of carbon-carbon bond length which
generates different phonon spectra at the $\Gamma $ point. Fig. 2 shows $%
\kappa (T)$ for three different bond lengths 1.42\r{A}, 1.44\r{A}\ and 1.46%
\r{A} for a zigzag (11,0) system. This result clearly shows that the basic
shape of $\kappa (T)$ is not changed qualitatively by varying the bond
length. The magnitude of $\kappa (T)$ is however dependent on bond length as
expected: the longer the bond length the lower the thermal conductance value.

In a real SWNT system the flow of thermal current is not perfect due to
scattering of phonons by various lattice imperfections. It is thus
inevitable that $\varsigma _{s}<1$ in general and only at resonances\cite
{ref23} do we have $\varsigma _{s}=1$ [see Eq. (7)]. Nevertheless, as long
as the transmission resonances are sharp for a large number of modes, which
is the case shown by previous numerical calculations for various
one-dimensional wires\cite{ref23}, we do not expect qualitative changes of
behavior from that shown in Fig. 2. On the other hand, it is interesting to
examine the opposite situation, namely the transmission is dominated by a
single resonance, {\it i.e.} \ the transmission coefficient $\varsigma _{s}$
is centered at a single mode $E_{0}$ with a resonance width $D$ [see
Eq.(7)]. Therefore for a small $D$ only very few terms in the summation of
Eq.(6) have substantial contribution to $\kappa (T)$, while a large $D$
should give a similar result as those of Fig. 2. Fig. 3 shows this behavior
for a number of values of $D$ by fixing $E_{0}=D$ for simplicity. Indeed, a
large $D$ which essentially gives $\varsigma _{s}=1$ for a wide range of
modes, gives a parabolic temperature dependence for $\kappa (T)$. For small
values of $D$ , on the other hand, not only the value of $\kappa $ is
reduced because only a few modes contribute, but also the temperature
dependence changes qualitatively. This exercise indicates the importance of
the quality of phonon wave transmission.

Recently the thermal conductance of the carbon nanotubes has been
experimentally measured\cite{ref6,ref7}. In both experiments, the results
show clearly that the tube-tube interactions are quite weak, implying that
it would be reasonable at some extent that is also capable of comparing the
measured results for multi-wall carbon nanotubes with the calculated ones
for single wall tubes without generating qualitative deviations, as it
indeed is. Our result is found to be qualitatively consistent with the
experimental data \cite{ref6} as shown in Fig. 4 where the circles are the
measured data and the dotted and dashed lines are our calculated results.
The solid line is a least square fit to the experimental data in the range
of $10K$ to $120K$ which gives $\kappa =$ $0.00089\ast
T^{1.97}(WK^{-1}M^{-1})$. Hence the experimental data indicate a parabolic
dependence on temperature. Surprisingly, our calculated value of $\kappa $
is even quantitatively reasonable considering the various approximation used
in our analysis: since the Lorentz number $\frac{k_{B}^{2}\pi ^{2}}{3h}\sim
10^{-12}(JK^{-2}S^{-1})$, the size of nanotube is on the order of $nm\sim
10^{-9}(M),$ and at $T=100K$ our calculated $\kappa \sim 100$ in units of
Lorenz number ({\it e.g}. Fig. 2), gives the same order of magnitudes in
magnitude as the measured data. To plot our calculated curves in Fig. 4, we
have used one experimental data point at $T=120K$ to fix the scale of $%
\kappa $. It is clear that for both zigzag and armchair tubes, the predicted
curves qualitatively agree with the measured data quite reasonably. \ In
Ref. \cite{ref7} , the measured $\kappa (T)$ is linear below 30K and an
upward bend slightly near 30K, i.e. the curve appears also to be
parabolic-like from 8K to 100K. 
The linear dependence below 30K in Ref.\cite{ref7} may be the results of
particular phonon scattering mechanisms or the imperfection phonon
transmission. 
As a result, our results appear to suggest that the thermal transport in
nanotubes might primarily be mediated by lattice vibrations, and the
electron contribution is less important.

\section{Summary}

We have studied the properties of the thermal conductance of single-walled
carbon nanotubes by means of a Landauer-Buttiker-like thermal conductance
formalism using a modified Tersoff-Brenner potential. We found that the
thermal conductance of the SWCN increases almost parabolically with
increasing temperature at perfect transmission at low temperature. Our
results are qualitatively consistent and quantitatively on the same order of
magnitude as the measured data\cite{ref6,ref7} because the tube-tube
interactions in these carbon nanotubes are weak. This allows us to suggest
that in the experimental device the phonon transmission is quite ideal,
which is consistent with the fact that the high quality of the nanotubes\cite
{ref6,ref7} was ensured in the measurements. It also suggests that it is the
phonon contribution which gave, to a large extent, the observed behavior of
temperature dependence of the thermal transport in SWCNs. In addition, our
results suggest that the anharmonic contribution to the thermal conductance
of SWCN is only quantitative but not qualitative.

Although the simple analysis presented in this paper has given a qualitative
understanding of the experimental data of Ref.\cite{ref6,ref7}, a number of
further improvements to theory are desirable. An extremely difficult subject
is the calculation of phonon transmission function for carbon nanotubes. Our
choice in this work is the phenomenological Breit-Wigner formula motivated
by one-dimensional theory of linear elasticity. For realistic nanotube
devices, especially those with lattice imperfections, it will be useful to
accurately compute the transmission coefficient. Another improvement to the
present work involves a more accurate analysis of the phonon frequency,
perhaps from first principle methods. Finally, since experimental
measurements are usually performed for a bundle of carbon nanotubes, if the
tube-tube interaction is not weak it is also interesting to investigate the
effect of inclusion of tube-tube interactions\cite{ref27}.

\acknowledgements 
One of authors (Q.R.Z.) is grateful to Prof. S.S. Xie concerning the
experimental data obtained by his lab. We gratefully acknowledge financial
support from the NSF of China (Grant No. 90103023 and 10104015), State Key
Project for fundamental Research of China and the Chinese Academy of
Sciences (G.S. and Q.R.Z.), RGC grant(HKU 7215/99P) from the Hong Kong
SAR(J.W.), and NSERC of Canada and FCAR of Quebec (H.G.).

\newpage \noindent {FIGURE CAPTIONS}\newline

\noindent Fig.1. The carbon-carbon two-body empirical potential as a
function of the bond length for four sets of parameters. Our numerical
analysis is based \ on the modified Brenner II potential.

\noindent Fig.2. Thermal conductance $\kappa $ versus temperature for
different C-C bond lengths for zigzag (11,0) SWCN assuming perfect phonon
transmission. Inset: The temperature-dependence of the thermal conductance
for armchair (11,11) and zigzag (11,0).

\noindent Fig.3. Thermal conductance $\kappa $ versus temperature for
different resonance width ($D$) fixing $D=E_{0}$ in unit of $cm^{-1}$.

\noindent Fig.4. Comparison of theoretical and experimental results.


\begin{references}
\bibitem{ref26}  The original Brenner$\prime $s parameters can also be used,
but we have checked that the first set gives smaller limiting frequency
while the second set gives a slightly larger limiting frequency.
\end{references}
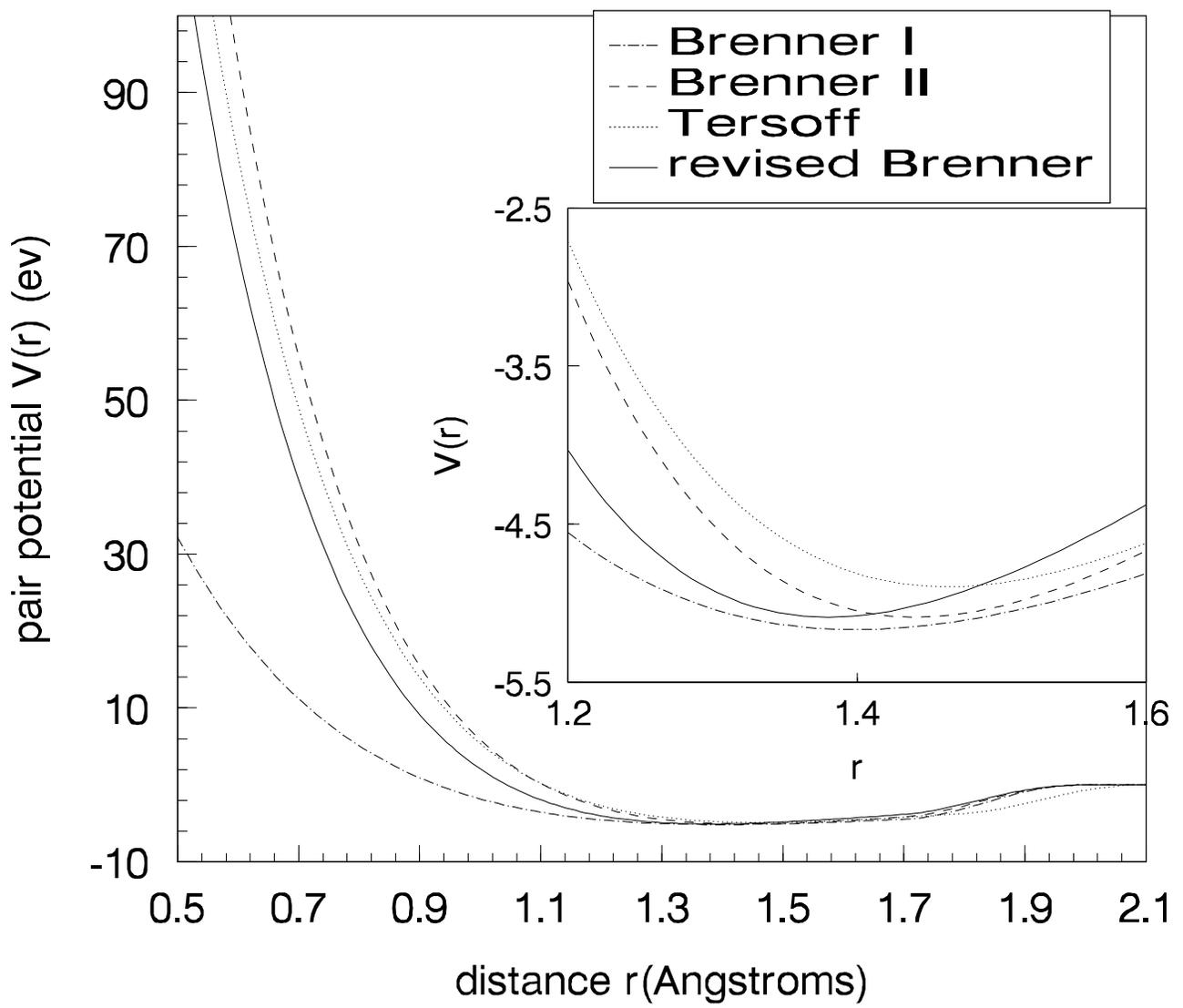
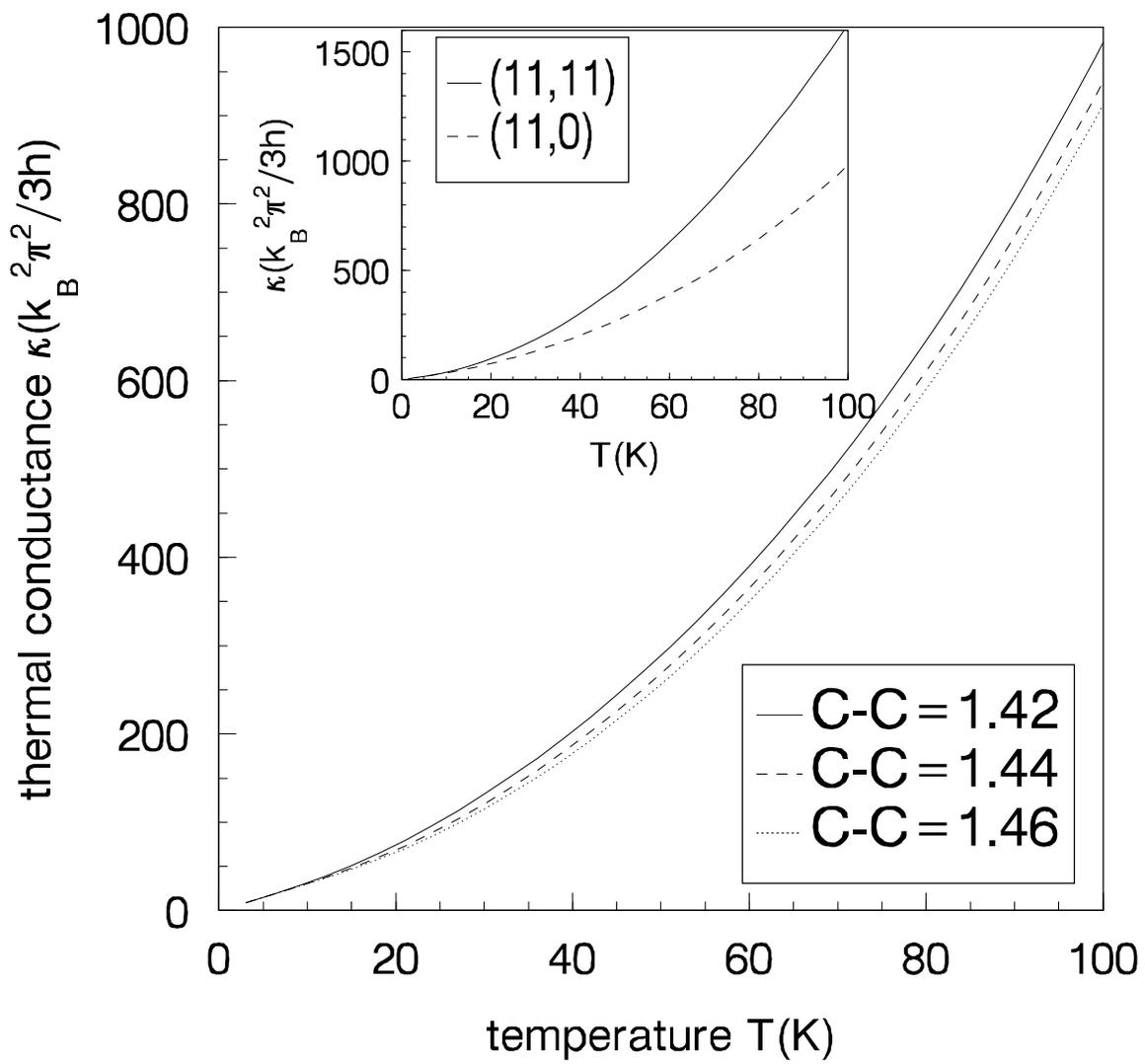
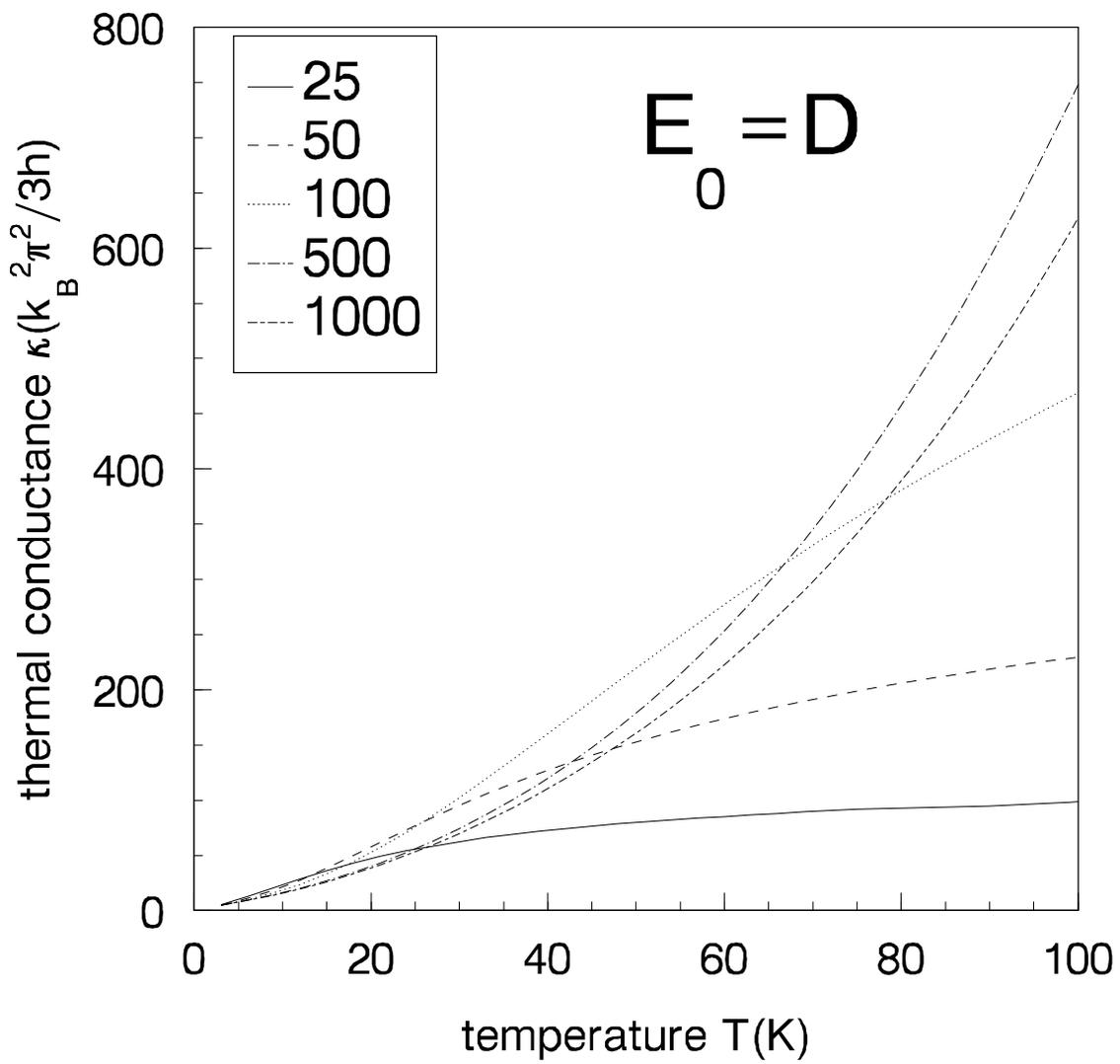
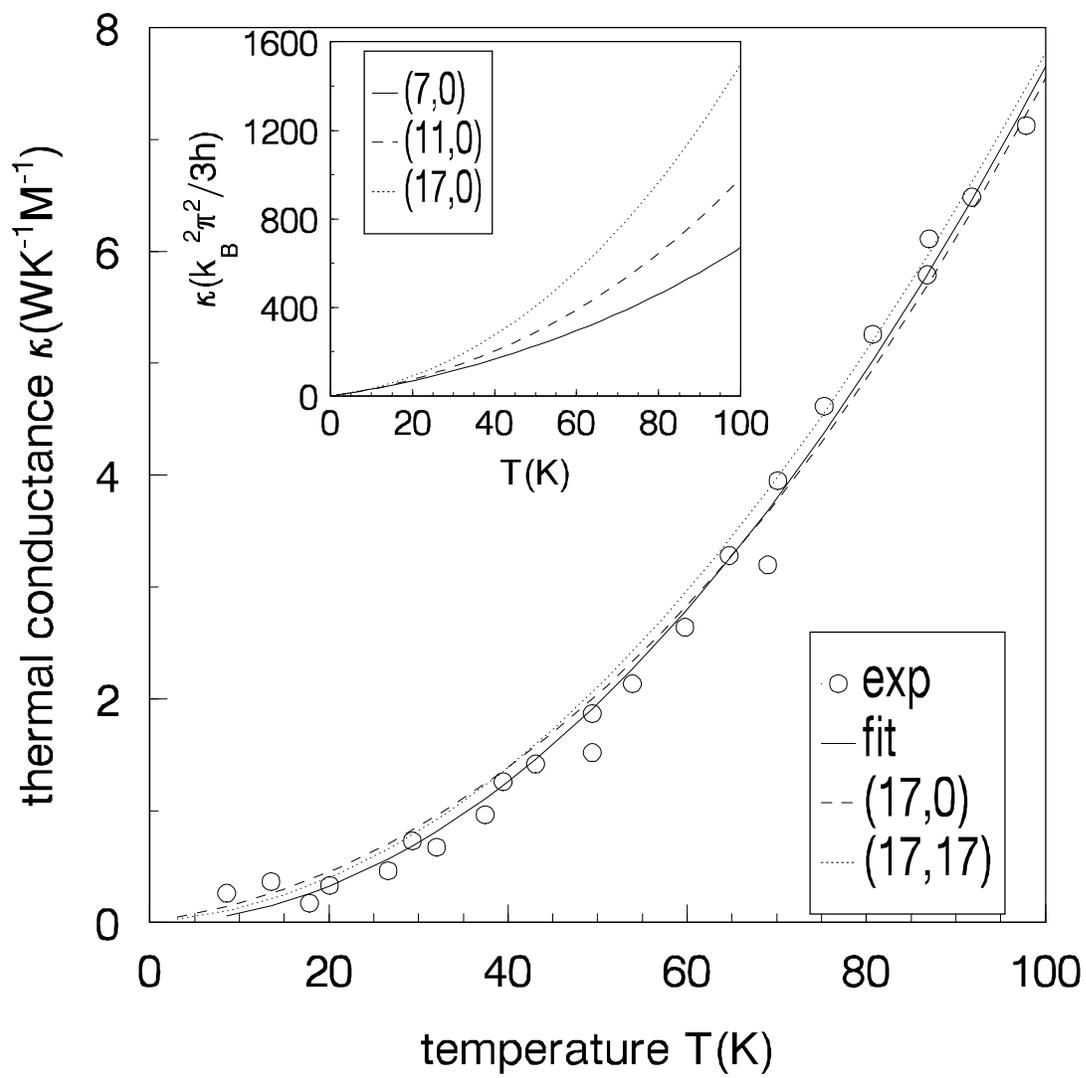
\end{document}